\documentclass[aps,prc,singspace,twocolumn,showpacs,groupaddress,superscriptaddress,floatfix,10pt,amsmath,amssymb,amsfonts]{revtex4-1}

\pdfoutput=1

\usepackage{graphicx}
\usepackage{wasysym}
\usepackage{subfigure}
\usepackage{array}
\usepackage{booktabs}
\usepackage{xfrac}
\usepackage{dcolumn}
\usepackage{color}

\begin{document}

\title{Limits on Tensor Coupling from Neutron $\beta$-Decay}
\author{R. W. Pattie Jr.}
\affiliation{Department of Physics, North Carolina State University, Raleigh, North Carolina 27695, USA}
\affiliation{Triangle Universities Nuclear Laboratory, Durham, North Carolina 27708, USA}
\email[Electronic address : ]{rwpattie@ncsu.edu}
\author{K. P. Hickerson}
\affiliation{Department of Physics and Astronomy, University of California at Los Angeles, Los Angeles, Calfornia, 90095, USA}
\author{A. R. Young}
\affiliation{Department of Physics, North Carolina State University, Raleigh, North Carolina 27695, USA}
\affiliation{Triangle Universities Nuclear Laboratory, Durham, North Carolina 27708, USA}

\date{\today}

\newcommand{\eref}[1]{Eq.~\ref{#1}}
\newcommand{\fref}[1]{Figure~\ref{#1}}
\newcommand{\tref}[1]{Table~\ref{#1}}
\newcommand{\aref}[1]{Appendix--\ref{#1}}
\newcommand{\sref}[1]{Section~\ref{#1}}
\newcommand{\unit}[2]{#1$^{#2}$}
\newcommand{\iso}[2]{$^{#2}$#1}
\newcommand{\CS}[0]{$C_\text{S}/C_\text{V}\;$}
\newcommand\T{\rule{0pt}{2.6ex}}
\newcommand\B{\rule[-1.2ex]{0pt}{0pt}}

\begin{abstract}
 Limits on the tensor couplings generating a Fierz interference term, b, in mixed Gamow-Teller Fermi decays can be derived by combining data from measurements of angular correlation parameters in neutron decay, the neutron lifetime, and $G_{\text{V}}=G_{\text{F}} V_{ud}$ as extracted from measurements of the $\mathcal{F}t$ values from the $0^{+} \to 0^{+}$ superallowed decays dataset.  These limits are derived by comparing the neutron $\beta$-decay rate as predicted in the standard model with the measured decay rate while allowing for the existence of beyond the standard model couplings.  We analyze limits derived from the electron-neutrino asymmetry, $a$, or the beta-asymmetry, $A$, finding that the most stringent limits for $C_\text{T}/C_\text{A}$ under the assumption of no right-handed currents is $-0.0026 < C_\text{T}/C_\text{A} < 0.0024$ (95\% C.L.) for the two most recent values of $A$.
\end{abstract}


\maketitle

\section{Introduction}

In the following we present an analysis of a ``lifetime-consistency test'' for neutron beta-decay, from which we derive relevant limits for beyond the standard model physics, in particular for new scalar and tensor couplings.  Our analysis utilizes high precision data from $0^{+}\rightarrow 0^{+}$ decays and neutron decay and does not supplant more general fitting procedure to obtain limits from all beta-decay data\cite{Sever2006,Wauters2013}. We note, however, that our limits are comparable to those obtained from fits to the entire beta decay set when similar assumptions are made (no right-handed neutrinos).  This brief report was inspired by comments in Bhattacharya et al. \cite{Bhattacharya2012} and begun as a part of thesis research \cite{Pattie2012}; however, we note that additional details have subsequently been published by Ivanov, Pitschmann, and Troitskaya \cite{Ivanov2012}.
 
The general method compares the measured value of the neutron lifetime, whose current Particle Data Group (PDG) value is $\tau_n = 880.1 \pm 1.1$ s, to a prediction of the neutron lifetime using the measured weak coupling strength from $0^+\to 0^+$ and the value of the axial-vector coupling constant,
$\lambda \equiv g_\text{A} /g_\text{V}$, extracted from angular correlations measurements.  Because this comparison requires the interpretation of specific angular correlations measurements to consistently extract limits, we analyze some specific cases of interest associated with the electron-neutrino correlation, $a$, and the beta-asymmetry, $A$.  We understand that this treatment is not exhaustive, nor should it supplant direct limits on the Fierz term in the neutron system, but it is intended to indicate the utility of these limits.

\section{Derivation of Impact of the Fierz Term on the Neutron Decay Rate}

$\beta$-decay can be represented, using all possible Lorentz-invariant couplings, by the Hamiltonian density
       \begin{eqnarray}
          \mathcal{H} & = & (\bar{p}n)(\bar{e}(C_{S} +
            C_{S}^{'}\gamma_5)\nu) \nonumber \\ &   & + (\bar{p}
            \gamma_\mu n)(\bar{e}\gamma_\mu (C_{V} +
            C_{V}^{'}\gamma_5)\nu) \nonumber \\ &   & + \frac{1}{2}
            (\bar{p}\sigma_{\lambda \mu} n)(\bar{e} \sigma_{\lambda
            \mu} (C_{T} + C_{T}^{'}\gamma_5)\nu) \nonumber \\ &   & -
            (\bar{p} \gamma_\mu \gamma_5 n)(\bar{e}\gamma_\mu \gamma_5
            (C_{A} + C_{A}^{'}\gamma_5)\nu) \nonumber \\ &   &
            +(\bar{p} \gamma_5 n)(\bar{e}\gamma_5 (C_{P} +
            C_{P}^{'}\gamma_5)\nu) + H.c. ,
       \end{eqnarray}
where $\sigma_{\lambda \mu} = -i/2(\gamma_\lambda \gamma_\mu - \gamma_\mu \gamma_\lambda)$ and p, n, e, and $\nu$ represent the hadronic and leptonic fields \cite{jackson1957}.  The strength of each type of interaction in the lepton current is given by a coupling
constant $C_{i}$ and $C_{i}^{'}$ where $i \in \{V,A,S,T,P\}$ are the vector, axial-vector, scalar, tensor, and pseudo-scalar interactions, respectively.
In the scenario where $\rvert C_{i} \rvert = \rvert C_{i}^{'} \rvert$, parity is maximally violated, and in the standard model  $\rvert C_{V}
\rvert = \rvert C_{V}^{'} \rvert$ and $\rvert C_{A} \rvert = \rvert C_{A}^{'} \rvert$ and $C_{S} = C_{S}^{'} =C_{T} = C_{T}^{'}=C_{P} =
C_{P}^{'} = 0 $.  These restrictions are experimentally determined, leaving the possibility for deviations below the current experimental
precision.

Limits on tensor couplings can be derived by noting that the decay rate for neutron $\beta$-decay can be written as (ignoring, at present, the possibility of a Fierz term )
\begin{equation}
\frac{1}{\tau_n} = \frac{G_{\text{V}}^2}{2 \pi^3 \hbar}(1+3\lambda^2) f_n  (1+\Delta_{\text{RC}}),
\label{eq:neutronraw}
\end{equation}
where, under the conserved vector current hypothesis, $G_\text{V} = G_\text{F} |V_{ud}|$, $f_n$ is the statistical rate function for the neutron defined as
\begin{align}
 f_n &= I_0(x_0)(1+\Delta_f) = 1.6887, \\
 I_k(x_0) &= \int^{x_0}_1x^{1-k}(x_0-x)^2\sqrt{x^2-1} \, \mathrm{d}x ,
 \label{eq:ratefunc}
\end{align}
and where $x$ and $x_0$ are the electron total energy and end-point energy in terms of the electron rest mass, and $\Delta_f$ is the Coulomb and recoil correction for the phase-space integral $I_0(x_0)=1.6299$. The standard model electroweak radiative corrections are denoted by $\Delta_{\text{RC}}=3.90(8)\times10^{-2}$ \cite{Czarnecki2004}. $G_{\text{F}}$ is the Fermi coupling constant as extracted from muon decay \cite{Profumo2007},  and $V_{ud}$ is the first element of the Cabibbo-Kobyashi-Maskawa (CKM) quark mixing matrix.  One can also predict the neutron decay rate from $0^+\to 0^+$ decays, by using the extracted value of $\tilde{G}_{\text{V}}$ from the average $\mathcal{F}t_{0^+\to0^+}$ and $\lambda$ from neutron angular correlation measurements, where if the Fierz term is zero $\tilde{G}_\text{V}^2 = G_\text{V}^2$
\begin{equation}
\frac{1}{\tau_{0^+}} = \frac{\tilde{G}_{\text{V}}^2}{2\pi^3 \hbar}(1+3\lambda^2)f_n (1+\Delta_{\text{RC}}).
\label{eq:zeropraw}
\end{equation}

A non-zero Fierz term will alter the neutron decay rate, $\tau_n$, via a $\langle m_e/E \rangle$ term in the phase-space integral and modify the value of $G_\text{V}$ extracted from superallowed Fermi decays to
\begin{equation}
 \tilde{G}_\text{V}^2  = G_\text{V}^2  \left\langle 1+b_F \gamma \frac{I_1(\tilde{x}_0)}{I_0(\tilde{x}_0)} \right\rangle,
 \label{eq:gvmeas}
\end{equation}
where $\gamma = \sqrt{1-(Z \alpha)^2}$, $Z$ is the atomic number, $\alpha$ is the fine structure constant, and $\tilde{x}_0$ is the end point energy for the superallowed Fermi decay isotopes, and $I_1(\tilde{x}_0)/I_0(\tilde{x}_0)$ corresponds to the ratio of phase-space integrals over the superallowed decay used in the determination of $V_{ud}$ and $b_F = 2\, \text{Re}\left( C_\text{S} / C_\text{V}\right)$ \cite{Hardy2009}.  For the moment we will ignore the changes in $\lambda$ induced by $b$, this will be addressed in the following sections.   In \tref{tab:phases}, the 13 isotopes included in the determination of
the average $\mathcal{F}t$ are listed with the absolute uncertainty on the measurement and the statistical rate function and the ratio $I_1(\tilde{x}_0)/I_0(\tilde{x}_0)$ \cite{Hardy2009}.  The reported values include both recoil and Coulomb corrections. Writing \eref{eq:neutronraw} and \eref{eq:zeropraw} in terms of $G_\text{V}$, $b_F$ and $b$, we have
\begin{equation}
\frac{1}{\tau_n} = \frac{G_{\text{V}}^2 }{2\pi^3 \hbar } (1+3\lambda^2)  f_n (1+\Delta_\text{RC})\left(1+\kappa b\right),
\label{eq:taun}
\end{equation}
and
\begin{equation}
  \frac{1}{\tau_{0^{+}}} = \frac{G_{\text{V}}^2}{2\pi^3 \hbar}(1+3\lambda ^2) f_n (1+\Delta_\text{RC}) \left(1 + \zeta b_F\right),
  \label{eq:tau0}
\end{equation}
where $\kappa = I_1(x_0)/I_0(x_0)$, $\tilde{\kappa} = I_1(\tilde{x}_0)/I_0(\tilde{x}_1)$, and $\zeta = \langle \gamma \tilde{\kappa}\rangle \sim 0.2560$. In \eref{eq:taun}, the term $(1+\kappa b)$ arises from the neutron phase-space integral when $b\ne0$, and the $(1 + \zeta b_F)$ term in \eref{eq:tau0} is from substitution of measured $G_\text{V}$ using \eref{eq:gvmeas}. Taking the  difference between the measured neutron decay rate and the decay rate predicted from $0^+\to0^+$ decays in terms of measured quantities gives
\begin{align}
   \tau_n K (1+3\lambda^2) = \frac{  1+ \zeta b_F}{1+ \kappa b }
   \label{eq:bdiffs}
\end{align}
where all the constants have been combined into \[K = \frac{\tilde{G}_{\text{V}}^2 f_n(1+\Delta_\text{RC})}{2\pi^3 \hbar} = 1.934(2)\times10^{-4} \,\text{s}^{-1},\] we express \eref{eq:bdiffs} in terms of the measured vlue of the weak coupling constant $\tilde{G}_\text{V}$ using \eref{eq:gvmeas}, and we are neglecting any affect on $\lambda$ due to $b$.  Critically, leading order differences in the predicted versus measured decay rates must come from scalar and tensor-induced couplings in the Fierz term, and any new physics which adjusts the value of $G_\text{F}$ and $V_{ud}$ affects both rates uniformly.  Additionally, the impact of the scalar coupling determined in the superallowed decays is suppressed by $\zeta$ due to the much higher endpoint energy of these decays relative to neutron $\beta$-decay.

\begingroup
\squeezetable
\begin{table*}
 \centering
  \caption{\label{tab:phases} The statistical weighting and ratio of the phase-space factors is presented for each of the 13 isotopes used in the $0^+\to0^+$ superallowed dataset to calculate the average $\mathcal{F}t$ value. $ I_k(\tilde{x}_0)$ are the statistical rate functions defined in \eref{eq:ratefunc} and calculated by Towner and Hardy \cite{Hardy2009,Townerpc13}.}
  \begin{ruledtabular}
  \begin{tabular}{c|dddd}
    Isotope \T \B & \mathcal{F}t & I_0(\tilde{x}_0) & I_1(\tilde{x}_0) & \multicolumn{1}{c}{$I_1(\tilde{x}_0)/I_0(\tilde{x}_0)$} \\
   \hline
      \iso{C}{10}\T   & 3067.7(4.6)  & 2.3004(12)   & 1.42401(74)    & 0.6190(5) \\
      \iso{O}{14}     & 3071.5(3.3)  & 42.772(23)   & 18.743(10)     & 0.4382(3) \\
      \iso{Mg}{22}    & 3078.0(7.4)  & 418.39(17)   & 128.948(52)    & 0.3082(2) \\
      \iso{Ar}{34}    & 3069.6(8.5)  & 3414.5(1.5)  & 724.56(32)     & 0.2122(1) \\
      \iso{Al}{26}$^m$& 3072.4(1.4)  & 478.237(38)  & 143.662(11)    & 0.3004(1) \\
      \iso{Cl}{34}    & 3070.6(2.1)  & 1995.96(47)  & 466.26(11)     & 0.2336(1) \\
      \iso{K}{38}$^m$ & 3072.5(2.4)  & 3297.88(34)  & 701.459(69)    & 0.2127(1) \\
      \iso{Sc}{42}    & 3072.4(2.7)  & 4472.24(1.15)& 895.34(23)     & 0.2002(1) \\
      \iso{V}{46}     & 3073.3(2.7)  & 7209.47(90)  & 1317.17(16)    & 0.1827(1) \\
      \iso{Mn}{50}    & 3070.9(2.8)  & 10745.97(57) & 1816.07(10)    & 0.1690(1) \\
      \iso{Co}{54}    & 3069.9(3.3)  & 15766.6(2.9) & 2470.63(45)    & 0.1567(1) \\
      \iso{Ga}{62}    & 3071.5(7.2)  & 26400.2(8.3) & 3719.7(1.2)    & 0.1409(1) \\
      \iso{Rb}{74}\B  & 3078.0(13.0) & 47300.0(110) & 5884.1(1.4)    & 0.1244(4) \\ 
  \hline 
     Average \T \B  & 3072.08(79)  &  &  & 0.2579(1) \\
  \end{tabular}
  \end{ruledtabular}
\end{table*}
\endgroup

Under of the assumptions of this analysis the Fierz inference term in neutron $\beta$-decay can be approximated in terms of the scalar \CS and tensor $C_\text{T}/C_\text{A}$ couplings \cite{Sever2006}
\begin{equation}
 b =  \frac{ 2 \sqrt{1-\alpha^2}}{1+3\lambda^2}  \bigg[ \text{Re} \bigg( \frac{C_{\text{S}}}{C_{\text{V}}}\bigg) + 3 \lambda^2 \text{Re} \bigg( \frac{C_{\text{T}}}{C_{\text{A}}} \bigg) \bigg].
 \label{eq:bdef}
\end{equation}
At this point, we already have a reasonably strong constraint on new physics by using \eref{eq:bdiffs}, \eref{eq:bdef}, and the definition of $b_F$
\begin{align}
 \frac{C_\text{T}}{C_\text{A}} (6\lambda^2 \gamma) &= \frac{\delta b}{\tau_n K \kappa } - 2 \gamma \frac{C_\text{S}}{C_\text{V}}  \nonumber \\
 &- (1+3\lambda^2)\left(\frac{\delta b}{\kappa} - 2 \zeta \frac{C_\text{S}}{C_\text{V}}\right),
 \label{eq:ctlamb}
\end{align}
where $\delta b = \left\langle 1+2 (C_\text{S}/C_\text{V})  \tilde{\kappa} \right\rangle$.  Using the PDG values for $\lambda = -1.2701 (25)$, $\tau_n = 880.1(1.1)$~s, and $V_{ud} = 0.97425(40)$ \cite{Beringer2012} and limits on scalar couplings, $C_\text{S}/C_\text{V} = 0.0011(13)$, from the superallowed dataset \cite{Hardy2009}, one can place a limit on the tensor coupling.  This results in 2-$\sigma$ (95\% C.L.) limits of $-0.0009 < C_\text{T}/C_\text{A} < 0.0125$.   Note that if only the Perkeo II result for $\lambda = -1.2739(19)$ \cite{Mund2012a} is used, then limits shift to $-0.0012 < C_\text{T}/C_\text{A} < 0.0065$.

\section{$b$ Dependence of $\lambda$}
The limits obtained from \eref{eq:ctlamb} have ignored the fact that $\lambda$ is determined experimentally by measuring correlation coefficients, which typically are modified by the existence of a Fierz term as in
\begin{equation}
 X_m(E_e) = \frac{X_0(E_e)}{1+b\, m_e/E_e},
 \label{eq:dilute}
\end{equation}
where $X_m(E_e)\in \{ a,A,B,... \}$ is the measured value of coefficient as a function of electron energy.  For this analysis we will focus on $a$ and $A$ since closed form expressions can be obtained for limits on tensor couplings and they have the highest sensitivity to $\lambda$. The method used to extract the correlation coefficient and the energy range of the analysis will impact the sensitivity to $b$, as will be shown explicitly in the case of $a$. Note: we are also explicitly ignoring imaginary couplings and the affect of tensor and scalar couplings on the angular correlations $a$ and $A$, which are second order in $C_\text{S}$ and $C_\text{T}$.

\subsection{$\lambda$ derived from the measured $\beta$-asymmetry parameter, A$_m$}
Single parameter fits to the energy dependence of the $\beta$-asymmetry will modify the measured quantity by $A_0/(1+b\langle I_1(x_1,x_2)/I_0(x_1,x_2)\rangle)$, where $x_1$ and $x_2$ are the limits of the energy range used in analysis. \tref{tab:bimpact} shows the ratio of the phase-space integrals over the reported analysis energy range for several of the experiments that measure the asymmetry and would be subject to the type of dilution analyzed here.  The leading order expression of $A_o$ (where one has already corrected the measured asymmetry for small radiative and recoil order corrections \cite{jackson1957}) in terms of $\lambda$ is
\begin{equation}
 A_0 = 2|\lambda| \frac{1-|\lambda|}{1+3\lambda^2},
\end{equation}
which, combined with \eref{eq:bdef}, \eref{eq:dilute}, and \eref{eq:ctlamb}, gives an relation for $C_\text{T}/C_\text{A}$ in terms of purely measured quantities
\begin{align}
 3 \lambda^2 &= \left[ \frac{h}{C_\text{T}/C_\text{A} \gamma + y} \right]\nonumber \\
           &= 3\left( \frac{ -1 - \sqrt{1 - A_m^2\left( 3 \tilde{C_\text{T}} + 2 / A_m \right) \tilde{C_\text{S}} }}{3 A_m \tilde{C_\text{T}}  + 2} \right)^2.
\label{equ:lambdasq}
\end{align}
In \eref{equ:lambdasq}, we have made the following substitutions $\tilde{C_x} = C_x \gamma +1$, $y=(\delta b/ \kappa) - 2 \zeta (C_\text{S}/C_\text{V})$, and $h=(\tau_n K)^{-1}-\gamma \kappa -y$. We assume here that the BSM scalar and tensor couplings make negligible contributions to the radiative and recoil-order corrections.  For BSM couplings at the $\approx$ 0.01 level, this should certainly be true, as can be seen by inspecting radiative corrections, which are precisely defined in references  \cite{Holstein1974,Gardner2001,Gudkov2006}.  \eref{equ:lambdasq} can be solved in closed form, producing three roots, two of which predict large values of $C_\text{T}/C_\text{A}$ and are ruled out by current experimental limits (see Appendix A).  The remaining solution gives a 2-$\sigma$ limit on the tensor coupling of $-0.0015 < C_\text{T}/C_\text{A}  < 0.0079 $, using the PDG values for the measured parameters. 

The PDG value of $\lambda$ includes the result of Mostovoi et al \cite{Mostovoi2001}, which is determined by simultaneous measurement of the $\beta$-asymmetry, $A$, and the neutrino asymmetry, $B$, and is therefore not consistent with the approach presented here.  Careful analysis of a simultaneous experiment would be required to determine the impact a non-zero Fierz term would have on the extracted $\lambda$.  Using the results from Perkeo II \cite{Mund2012a} and UCNA \cite{Mendenhall2013}, $A_m = -0.11931(46)$, we then obtain $-0.0026 < C_\text{T}/C_\text{A} < 0.0024$ (95\% C.L.).  Note that the 30\% reduction in the limit is due in large part to the increased error bar on $A_0$ used by the PDG to account for the variations in the current measurements; including this factor would increase the limit to $2\sigma = 0.0039$. Future prospects for reducing this limit via increasing the precision of $A_m$ are shown in \fref{fig:ctvsla}, where we see that next generation experiments will reduce the uncertainty on $A_m$ to the point where $\delta \tau_n$ becomes the leading contribution to this limit.

\begingroup
\squeezetable
\begin{table*}
 \centering
 \caption{\label{tab:exprs}Experimental results for $\lambda$ from measurements of $A$ using a single parameter fit to energy dependence of the asymmetry are summarized. For each measurement the reported analysis window and the phase-space integral ratios over that range are listed. The last column estimates the change in the asymmetry where $b=0.001$.}
\begin{ruledtabular}
 \begin{tabular}{l|llc|cc|c}
   Experiment                            &  A                & $\lambda$    & Energy Range & $I_0(x_1,x_2)/I_0(x_0)$ & $I_1(x_1,x_2)/I_0(x_0)$ &$\Delta A$ [\%]\\
  \hline 
   Perkeo \cite{Bopp1986} \T             & -0.1146(19)       & -1.262(5)       &   $>$200 keV  &  0.801          & 0.581        &   0.06   \\
   Perkeo II \cite{Mund2012a}            & -0.11951(50)      & -1.2755(13)     &   325-675 keV &  0.843          & 0.534        &   0.05   \\
   Ill TPC \cite{Liaud199753}	         & -0.1160(9)(12)    & -1.266(4)       &   200-700keV  &  0.807          & 0.583        &   0.06   \\
   UCNA \cite{Mendenhall2013}  	         & -0.11952(110)  & -1.2756(30)     &   275-625 keV &  0.828          & 0.557        &   0.06   \\
   Yerozolimsky \cite{Yerozolimsky1997240}\B    & -0.1135(14)       & -1.2594(38)     &   250-780 keV &  0.824          & 0.561        &   0.06   \\
 \end{tabular}
 \label{tab:bimpact}
\end{ruledtabular}
\end{table*}
\endgroup

\subsection{$\lambda$ derived from the electron-neutrino correlation parameter $a_0$}

\begin{figure}[hb!]
 \centering
 \includegraphics[width=0.98\linewidth]{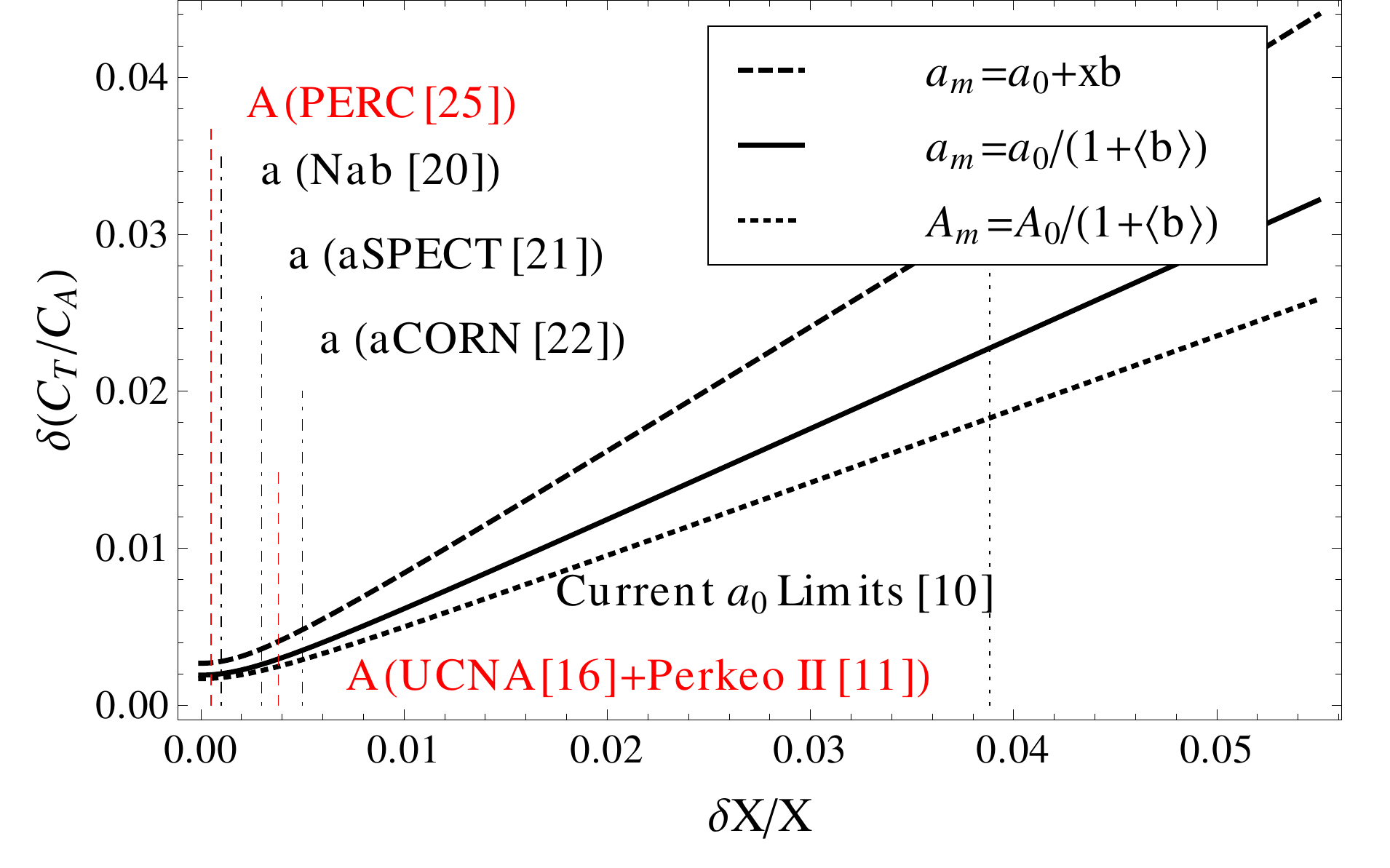}
 \caption{\label{fig:ctvsla}The 2-$\sigma$ limits on $C_\text{T}/C_\text{A}$ are shown for the methods of extracting $a$ discussed in the text, along with those from fitting the energy dependence of the $\beta$-asymmetry $A$.  Current limits on $a$ are taken from the PDG2012 \cite{Beringer2012}, and future limits denote the proposed sensitivity of experiments such as Nab\cite{Pocanic2009}(directional method), aCORN \cite{Wietfeldt2009}(directional method), and aSPECT \cite{Baesler2008}(fit method).  For this analysis we consider the $\beta$-asymmetry from UCNA \cite{Mendenhall2013} and Perkeo II \cite{Mund2012a}, and the proposed limits are from the  PERC experiment \cite{Dubbers2008}.}
\end{figure}

Measurements of the electron-neutrino correlation parameter $a_0$ are being proposed and carried out at both cold and ultracold neutron facilities world wide with the aim of significantly improving the current precision of 3.9\% to $<0.1\%$  \cite{Pocanic2009,Baesler2008,Wietfeldt2009}.  Determining the $a$ coefficient can be performed by directly measuring the angular distribution of emitted electron and proton in coincidence, in which case the $a_0/(1+b m_e/E_e)$ scaling can apply (directional method), or via a measurement of the proton energy spectrum.  Directly fitting the proton spectrum or using discrete points from the spectrum (spectral fit method) as in Stratowa et al \cite{Stratowa1978} will result in $a_m = a_0 + x_f b$, where $x_f \sim 0.09$ as determined by Monte Carlo. An alternate method of analyzing proton spectral data is an integral analysis, which has the advantage of being much less sensitive to the presence of a Fierz term, as presented in \cite{Gluck1995}.In such analysis one compares the integral rate over a fixed energy range to the total decay rate. This results in a linear scaling of the form
\begin{equation}
 a_m =a_0 +x_I b = \frac{1-\lambda^2}{1+3\lambda^2} + x_I b,
 \label{eq:adef}
\end{equation}
where $a_m=-0.103(4)$ is the measured value of the correlation coefficient and $x_I \sim 0.008$ from \cite{Gluck1995}, which was confirmed by this analysis using Monte Carlo.  Using \eref{eq:bdef}, \eref{eq:ctlamb}, and the two expressions for the measured coefficient including a Fierz term, we can find a solution for $C_\text{T}/C_\text{A}$ in terms of measurement quantities

\begin{equation}
 \frac{C_\text{T}}{C_\text{A}} = \left\{ 
    \begin{array}{l l}
     \frac{h\left(a+\frac{1}{3}\right)-y\left(1-a_m+x_I \gamma \frac{C_\text{S}}{C_\text{V}}\right)}{\gamma \left(1-a_m+x_I \gamma \frac{C_\text{S}}{C_\text{V}}\right)-x\gamma h} & \text{(Linear)}, \\
     \frac{h\left(1+\frac{1}{3 a_m}\right)-y\left(\frac{1}{a_m} - 1 - \gamma \kappa \frac{C_\text{S}}{C_\text{V}}\right)}{\gamma \left(\frac{1}{a_m} -1 -\kappa \gamma \frac{C_\text{S}}{C_\text{V}} - h \kappa\right)} & \text{(Inverse)},
    \end{array}
     \right. 
\end{equation}
where we have made use of the substitutions defined in the previous section.  This approach is more straightforward due to the fact that, unlike $A$, there are no terms which are linear in $\lambda$ in \eref{eq:adef}.  Using the current PDG value for $a_0=-0.1030(40)$ (extracted using the spectral fit method, $x_f = 0.09$)  we find limits of 
\begin{equation}
-0.0134 < \frac{C_\text{T}}{C_\text{A}} < 0.0324 \; (95\% C.L.).
\end{equation}
While this limit is not currently competitive with those obtained through measurements of the $\beta$-asymmetry, \fref{fig:ctvsla} shows that determining $a$ from angular distributions is more sensitive than the spectral fit method to tensor couplings, and the next generation experiments should improve upon current limits set by $A_0$.  (For example, a measurement such as Nab \cite{Pocanic2009}, aiming for a precision of $\Delta a/a \simeq 10^{-3}$, could set constraints of $|C_\text{T}/C_\text{A}| < 0.0015$ with the current uncertainty in the lifetime.)

We also note that the differing sensitivities to the Fierz term afforded by the integral proton spectrum analysis and directional distribution measurements afford an alternate method to extract limits on the Fierz term.

\section{Conclusion}
In this analysis we have presented a self-consistent derivation of limits to tensor couplings in the weak interaction, using experimentally measured quantities from neutron $\beta$-decay and $0^+\to0^+$ superallowed Fermi decays.  By calculating the difference of the measured and predicted neutron $\beta$-decay rate, we are able to derive a limit to the tensor coupling $-0.0026< C_\text{T}/C_\text{A} < 0.0024$,  under the assumption of maximal parity violation and no right handed neutrinos, where $C_X = C_X^{'}$.  Noting that a non-zero Fierz term would modify the experimentally reported value of $\lambda$, we have shown that the measured correlation coefficients $a_m$ and $A_m$ can be used to set limits on the tensor couplings that are competitive with those obtained from global fits to the available data \cite{Wauters2013}.  If the precision of $A$ or $a$ reaches the 0.1\% level, then the accuracy of the neutron lifetime becomes the leading contribution to the derived limits. 

These results can be used to set constraints on the effective couplings from Bhattacharya et al \cite{Bhattacharya2012}, where the tensor coupling is given as
\begin{equation}
 \frac{C_\text{T}}{C_\text{A}} = \frac{-4 g_T \epsilon_T}{g_\text{A}(1+\epsilon_L - \epsilon_R)},
\end{equation}
where $\epsilon_{R(L)}$ represent the effective right and left handed couplings and both are zero in the SM.  In general, this directly leads to a limit of $ -5.8\times10^{-4}< g_T\epsilon_T/(g_\text{A}(1+\epsilon_L- \epsilon_R) < 6.4\times10^{-4}$.  However, under the assumption that BSM physics arises from tensor couplings then $\epsilon_\text{R}=\epsilon_\text{L} = 0 $, and this simplifies to a limit on $g_\text{T}\epsilon_\text{T}/g_\text{A}$.  

\acknowledgements
We would like to thank V. Cirigliano and J.C. Hardy for many helpful discussions and I.S. Towner for providing calculations of $I_1/I_0$ for the superallowed dataset in \tref{tab:phases}.   This work was supported by NSF grant 1005233 and DOE grant number DE-FG02-97ER41042.

\appendix*
\section{Solutions for $C_\text{T}/C_\text{A}(A_m)$}

The roots obtained from solving \eref{equ:lambdasq} are 
\begin{widetext}
\begin{equation}
C_\text{T}/C_\text{A} = \left\{
  \begin{array}{l l}
   -\frac{2+3 A_m}{3 A_m \gamma} ,\\
   -\frac{ A_m (2+3 A_m) h^2 +3 y A_m(1+ C_s \gamma )^2+ h  [A_m \{2+3 A_m (1+y)\} (1+C_s \gamma )-2]\pm2 h\gamma \sqrt{ 1+A_m \{3 A_m (y-1)-2\} (1+h+C_s \gamma )}}{3 A_m^2 \gamma (1+h+C_s \gamma )^2},
  \end{array}
  \right.
\end{equation}
\end{widetext}
where $h$, $y$, and $\gamma$ are defined in the text and $C_\text{S} \equiv C_\text{S}/C_\text{V}$.  The limits obtained from the two roots not discussed in the text are 
\begin{equation}
 \frac{C_\text{T}}{C_\text{A}} = \left\{
 \begin{array}{l l}
  4.67 \pm 0.05 , (\textrm{first root})\\
  25.8 \pm 0.9 , (\textrm{negative root})
 \end{array}
 \right.
\end{equation}
both of which predict large central values for the tensor coupling that are $>25 \sigma$ from zero and can therefore be neglected as unphysical.

\bibliography{tensor_limits_paper}{}
\bibliographystyle{apsrev}

\end{document}